\documentclass[aps,pra,longbibliography,twocolumn,groupedaddress,amsmath,amssymb]{revtex4-1}
\usepackage{graphicx}  % needed for figures
\usepackage{dcolumn}   % needed for some tables
\usepackage{bm}        % for math
\usepackage{verbatim}   % for math
\usepackage{url}
\usepackage{xcolor}
\usepackage{subcaption}
\usepackage[colorlinks, linkcolor=blue,anchorcolor=blue,citecolor=blue,urlcolor=blue]{hyperref}
\begin{document}

\title{Coupling a single NV center with a superconducting qubit via the electro-optic effect}
\author{Chang-Hao Li}
\affiliation{
   Department of Applied Physics,
   Xi'an Jiaotong University,
   Xi'an 710049, China
   }
\author{Peng-Bo Li}
\email{lipengbo@mail.xjtu.edu.cn}
\affiliation{
   Department of Applied Physics,
   Xi'an Jiaotong University,
   Xi'an 710049, China
   }

\begin{abstract}
We propose an efficient scheme for transferring quantum states and generating entangled states between two qubits of different nature. The hybrid system consists a single nitrogen vacancy (NV) center and a superconducting (SC) qubit, which couple to an optical cavity and a microwave resonator, respectively. Meanwhile, the optical cavity and the microwave resonator are coupled via the electro-optic effect. By adjusting the relative parameters, we can achieve high fidelity quantum state transfer as well as highly entangled states between the NV center and the SC qubit.  This protocol is within the reach of currently available techniques, and  may provide interesting applications in quantum communication and computation with single NV centers and SC qubits.
 \end{abstract}

\maketitle

\section{Introduction}
Hybrid quantum systems that combine two or more subsystems can harness the strengths of different platforms and point a way toward future quantum technologies, including quantum detectors, simulators, and computers \cite{RevModPhys.85.623,RevModPhys.86.153}. Among these subsystems, superconducting (SC) circuits \cite{RevModPhys.73.357,10.1038/nature10122, 10.1038/nature07128,nature_SC,10.1038/nature02851,10.1038/nphys1730,PhysRevB.68.064509,PhysRevA.69.062320,yang2017entangling,li2012engineering} are considered one of the most promising platforms for quantum information processing. They can couple strongly to electromagnetic fields, which makes rapid logic gate operations possible. However, they are limited by  short coherence time due to high sensitivity to noises. On the other hand, spin systems such as nitrogen-vacancy (NV) centers in diamond have rather long coherence time even at room temperature \cite{Hanson352,10.1038/nmat2420,0953-8984-18-21-S08,Maurer1283,prl-117-015502,prl-116-043603}.  Therefore, recently much attention has been paid to the integration of SC circuits (quantum processors) and atoms or spins (quantum memory) \cite{RevModPhys.73.357,astner2017coherent,deng2015charge}.

Among many types of hybrid quantum systems, coupling NV centers in diamond with a superconducting qubit is quite appealing \cite{PhysRevLett.105.210501,PhysRevB.81.241202,PhysRevA.84.010301,PhysRevB.87.144516,PhysRevA.88.012329,pra-91-062325,pra-91-042307,Chen:15,PhysRevA.92.052335}. Strong coupling of superconducting qubits to NV ensembles has been proposed theoretically \cite{PhysRevLett.105.210501} and demonstrated experimentally \cite{10.1038/nature10462,PhysRevLett.107.220501}. However, the spin ensembles have much shorter coherence time resulted from inhomogeneous broadening \cite{PhysRevB.80.115202,PhysRevB.82.201201}. Therefore, it is desirable to couple a single NV center to a superconducting qubit via a quantum bus. Recently, proposals with quantum buses based on optomechanical and electromechanical couplings have attracted much attention \cite{PhysRevLett.105.220501,prappl-4-044003,PhysRevA.84.042341,QST_NVSC,Entangle_NVSC,PhysRevA.93.062336,PhysRevA.91.012333}. However, the prerequisite of very low temperatures and ground state cooling in those coupling schemes has posed other technical challenges \cite{10.1038/nature08093,10.1038/nature08967,10.1038/nature10261,10.1038/nature10461}. On the other hand, the  ground state for optical fields can be taken as the vacuum state even at ambient temperature  and it is much easier to control optical photons using quantum optical methods. At the same time, a single NV center and a superconducting qubit can strongly couple to an optical cavity \cite{10.1021/nl061342r,10.1021/nl8032944,10.1021/nl8023627,PhysRevA.83.054306,PhysRevA.85.042306,OE-25-16931,10.1038/nature09256} and a microwave (MW) resonator (such as coplanar waveguide and LC oscillators) respectively \cite{nature_SC,10.1038/nature02851,10.1038/nphys1730,PhysRevB.68.064509,PhysRevA.69.062320}. Therefore, it is much better to utilize a coherent optical-microwave interface as the quantum bus for coupling these two qubits of different nature.

In this work, we propose a scheme for efficiently transferring quantum states and generating entangled states between a single NV center and a superconducting qubit in a hybrid electro-optic system. The single NV center and the superconducting qubit are, respectively, coupled to an optical cavity and a microwave resonator, while the optical cavity and the microwave resonator are coupled via the electro-optic effect \cite{PhysRevA.81.063837,PhysRevA.84.043845,PhysRevA.94.053815}. It has been demonstrated that the electro-optic coupling has the same form as the optomechanical coupling via radiation pressure \cite{ Matsko:07}. Therefore, quantum effects in optomechanical schemes \cite{RevModPhys.86.1391} can in principle be observed in electro-optic systems.

We show that by applying a red-sideband laser driving, the optical cavity and the microwave resonator can interact with each other in the beam-splitter form. The simultaneous creation and annihilation of both Boson modes can lead to quantum state transfer between the two qubits. On the other hand, with a blue-sideband driving the electro-optic interaction can have a two-mode squeezing form, which is responsible for highly entangled states of the NV and SC qubits. In particular, we find both proposals work best in the intermediate coupling regime, where the electro-optic coupling strength is comparable to those in the NV-cavity and SC-resonator subsystems. We also consider the experimental feasibility of the proposal and show the parameters here are in line with current techniques. Our scheme is easy to control and does not require ultralow temperature and additional ground state cooling. This work may provide possible applications in quantum information processing based on hybrid systems consisting of a single spin qubit and a superconducting qubit.

\section{The setup}

\captionsetup{labelfont=bf,textfont=normalfont,singlelinecheck=off,justification=raggedright}
\begin{figure}
\includegraphics[scale=0.26]{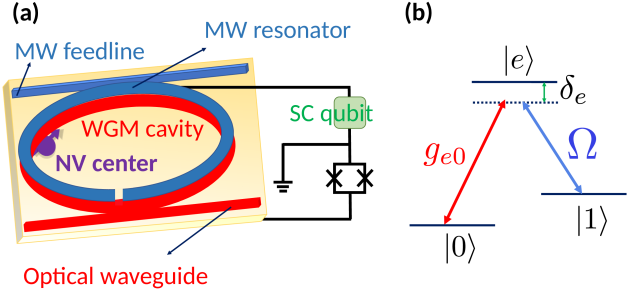}
\caption{\label{Exp_setup} (a) The schematic diagram of the proposal. An NV center and a superconducting qubit coupled to a WGM cavity and a MW resonator respectively. The electro-optic material (EOM) is embedded in silica of the WGM cavity (in red) and thus is clamped. The top-electrode (in blue) is overlapped with the optical microring thus maximizing the electro-optic coupling. Together with the ground electrode (in yellow), they form a capacitor and can be utilized in a quantum circuit (here we show the case for a charge qubit) to couple a superconducting qubit (in green). The optical cavity is coupled to a waveguide via evanescent field. (b) Effective energy-level configuration of the NV center. The cavity mode is coupled to the transition $|0\rangle \leftrightarrow |e\rangle$ with strength $g_{e0}$ and there is a classical field corresponding to $|1\rangle \leftrightarrow |e\rangle$ with strength $\Omega$.   }
\end{figure}

As illustrated in Fig. \ref{Exp_setup}, we consider two subsystems: a single NV center is fixed on the exterior surface of a whispering-gallery mode (WGM) micro-cavity and a SC qubit is coupled to a superconducting microwave resonator. Meanwhile the two subsystems are coupled via the electro-optic effect \cite{PhysRevA.81.063837,PhysRevA.84.043845}. Since the EOM is clamped in the cavity, its mechanical degree of freedom is frozen. Also note that to ensure that only the positive phase of the microwave electric field profile couples to the WGM cavity, the symmetry of the microwave resonator should be broken, as shown in Fig. \ref{Exp_setup}. This integrated on-chip device of WGM cavity and MW resonator coupling has been investigated in detail in \cite{PhysRevA.94.053815}.

We consider one optical mode and assume $\hbar=1$ hereafter. The Hamiltonian of the optical cavity can be written as $H_{a}=\omega_{a}a^{\dagger}a$, where $\omega_{a}$ is the cavity frequency and $a^{\dagger}$($a$) is the creation(annihilation) operator of the cavity mode. The optical cavity is driven by a laser field, which is characterized by its frequency $\omega_{L}$ and amplitude $E_0$, i.e., $H_d=i E_0 a e^{i\omega_Lt}+\text{H.c.}$

The NV center has a ground triplet state denoted as $|^{3}A_{2}\rangle = |E_{0}\rangle \otimes |m_{s}=0, \pm1 \rangle$ where $|E_{0}\rangle$ labels orbital state with zero angular momentum projection along the N-V axis and there exists a zero-field splitting $D_{g}=2.87$ GHz between $|m_{s}=0 \rangle$ and $|m_{s}=\pm1 \rangle$ states. After breaking the degeneracy of $|m_{s}=\pm1 \rangle$ states with an external magnetic field, we can get a $\Lambda-$level system consisting of two ground states labeled by $|0\rangle=|E_{0}\rangle \otimes |m_{s}=-1\rangle, |1\rangle=|E_{0}\rangle \otimes |m_{s}=1\rangle$ and an excited optical state $|e\rangle =|A_{2}\rangle=\frac{1}{\sqrt2}(|E_{-}\rangle|m_{s}=+1\rangle+|E_{+}\rangle|m_{s}=-1\rangle)$ where $|E_{\pm}\rangle$ are orbital states with angular momentum projection $\pm 1$ along N-V axis. In the limit of low strain, the $|A_{2}\rangle$ state is robust with the stable symmetric properties and decays with equal probability to the ground levels $|m_{s}=+1\rangle$ and $|m_{s}=-1\rangle$.

The optical cavity mode is coupled to the transition $|0\rangle \leftrightarrow |e\rangle$ with strength $g_{e0}$, and a classical field
is applied to the transition $|1\rangle \leftrightarrow |e\rangle$ with frequency  $\omega_\Omega$ and strength $\Omega$.  The detunings for these transitions $|0\rangle  \leftrightarrow |e\rangle $  and $|1\rangle  \leftrightarrow |e\rangle $ are $\delta_{e}=\omega_{e0}-\omega_{a}=\omega_{e1}-\omega_{\Omega}$, as denoted in Fig.\ref{Exp_setup}(b). Here $\omega_{e0}$ and
$\omega_{e1}$ are the optical transition frequencies for the corresponding transitions.
Under the rotating wave approximation, the Hamiltonian describing the coupling between the NV center and the cavity mode as well the classical field reads

\begin{eqnarray}
H_{1,int}=g_{e0}a|e\rangle \langle 0|+\Omega e^{i\omega_{\Omega} t} |e\rangle \langle1| +\text{H.c}.
%\right)
\end{eqnarray}

In the limit of strong laser driving, we can go into a displaced picture, in which the cavity bosonic operator is written as the sum of its steady-state value and a small linear displacement, i.e., $a \rightarrow \alpha_{a}+\delta a$. Here the displacement operators obey the Gaussian statistics and have delta correlation relations in time with the average $\langle \delta a \rangle =0$. Then in the displaced picture, we have the interaction Hamiltonian
\begin{eqnarray}
H_{1,int}&=&g_{e0}\alpha_ae^{i\Delta_Lt}|e\rangle \langle 0|+g_{e0}\delta ae^{i\delta_et}|e\rangle \langle 0|\\ \nonumber
&&+\Omega e^{i\delta_et} |e\rangle \langle1| +\text{H.c},
\end{eqnarray}
with $\Delta_L=\omega_{e0}-\omega_L$.
Then we consider the system under the large detuning conditions, i.e., $\{|\delta_{e}|, \vert \Delta_L\vert,  \vert \Delta_L-\delta_{e}\vert\}\gg \{ |\Omega|, |g_{e0}|, \vert g_{e0}\alpha_a\vert\}$. In this case, the excited state $|e\rangle$ can be adiabatically eliminated as it dispersively couples to the other two states. Note that the laser field can drive the NV spin after the linearization process.  But after we perform adiabatic elimination of the excited level, it only leads to a shift of energy levels due to the Stark effect, which can be eliminated through adding static fields.   We then obtain the effective Hamiltonian for the NV-cavity subsystem:

\begin{eqnarray}
H_{1,eff}=g_{1}(\delta a\sigma_{NV}^{\dagger}+\delta a^{\dagger}\sigma_{NV})
%\right)
\end{eqnarray}
with $g_{1}=\frac{\Omega g_{e0}}{\delta_{e}}$ and $\sigma_{NV}=|0\rangle_{NV} \langle 1|$.

The SC qubit can be treated as a two-level system which has a ground state $|0\rangle_{SC}$ and excited state $|1\rangle_{SC}$ with an energy splitting $\mu$. Similar to the optical cavity, the microwave resonator can be modeled as a single Bosonic mode with $H_{b}=\omega_{b}b^{\dagger}b$ where $\omega_{b}$ and $b$ are the mode frequency and the annihilation operator respectively. The corresponding Hamiltonian in the interaction picture can be written as \cite{PhysRevB.68.064509,PhysRevA.69.062320}:
\begin{eqnarray}
H_{2,int}=g_{2}(b^{\dagger}\sigma_{SC}^{-}+b\sigma_{SC}^{+})
%\right)
\end{eqnarray}
where $\sigma_{SC}^{+}=|1\rangle_{SC}\langle0|$ and $\sigma_{SC}^{-}=|0\rangle_{SC}\langle1|$ are the raising and lowering operators of the SC qubit.

Now we consider coupling the cavity optical field to the resonator microwave field via the electro-optic effect \cite{PhysRevA.81.063837}. The medium incorporated in the optical cavity is a transverse EOM which induces voltage-dependent shift to the optical field. The interaction Hamiltonian then has the form  $H_{i}=\frac{\phi}{\tau}a^{\dagger}a$, which is the same as the one for optomechanics. Here $\tau$ is the round-trip time of light in the EOM and $\phi$ is the round-trip phase shift. The shift is given by \cite{QuantumElectronics} $\phi=\frac{\omega_{a}n^{3}lc_{eo}}{cd}V$, where  $n$ is the refractive index of the modulator, $c_{eo}$ is the electro-optic coefficient,  $l$ and $d$ are the length and thickness of the medium respectively,  $c$ is the speed of light, and  V is the voltage across the EOM.

One can model the modulator as a capacitor of a single-mode microwave resonator, then the voltage $V$ can be quantized in the form of $V=(\frac{\omega_{b}}{2C})^{\frac{1}{2}}(b^{\dagger}+b)$, where $C$ is the capacitance of the microwave resonator. Therefore the interaction Hamiltonian can be written as \cite{PhysRevA.81.063837}:
\begin{eqnarray}
H_{i}=g_{i}(b^{\dagger}+b)a^{\dagger}a
\end{eqnarray}
\begin{eqnarray}
g_{i} \equiv \frac{\omega_{a}n^{3}lc_{eo}}{c\tau d}(\frac{\omega_{b}}{2C})^{\frac{1}{2}}.
\end{eqnarray}
Taking into account the effect of the driving laser, it is a good
approximation to linearize the above Hamiltonian through replacing the optical annihilation
operator $a$ with the sum of its stable mean value $\alpha_a$ and its fluctuation term $\delta a$.
Then we obtain a linear interaction between the optical cavity and the microwave resonator:
\begin{eqnarray}
H_{linear}=G_{i}(b^{\dagger}+b)(\delta a^{\dagger}+\delta a)
\end{eqnarray}
where the coupling strength $G_{i}=g_{i}|\alpha_{a}|$ can be enhanced by the laser driving.

Now, the total system can be described by the following effective Hamiltonian:
\begin{eqnarray}
\begin{split}
&H_{total}=H_{1,eff}+H_{2,int}+H_{linear} \\
&=g_{1}(\delta a\sigma_{NV}^{\dagger}+\delta a^{\dagger}\sigma_{NV})+g_{2}(b^{\dagger}\sigma_{SC}^{-}+b\sigma_{SC}^{+}) \\
&+G_{i}(b^{\dagger}e^{i\omega_bt}+be^{-i\omega_bt})(\delta a^{\dagger}e^{i\Delta t}+\delta ae^{-i\Delta t}), \\
\label{H_total}
\end{split}
\end{eqnarray}
with $\Delta=\omega_a-\omega_L$.
The last term of the above Hamiltonian indicates the linear interaction between the optical cavity and the microwave resonator. Therefore, the interaction between the NV center and the superconducting qubit can be dynamically manipulated. In the following, we will show that with different sideband laser driving, we can achieve high fidelity quantum state transfer as well as highly entangled states between the NV center and the superconducting qubit.

%-----------------------------------------------------------------------------------------------------------------------------------------------------------------
\section{High fidelity quantum state transfer}
We now consider the case where the laser driving is in the red-sideband, i.e., $\Delta=\omega_a-\omega_L=\omega_{b}$. Under the rotating-wave approximation, from Eq.(\ref{H_total}) we can obtain the Hamiltonian in the interaction picture
\begin{eqnarray}
\begin{split}
&H_{transfer}=g_{1}(\delta a^{\dagger}\sigma_{1}+\delta a\sigma_{1}^{+})+g_{2}(b^{\dagger}\sigma_{2}+b\sigma_{2}^{+}) \\
&+G_{i}(\delta a^{\dagger}b+b^{\dagger}\delta a),
\end{split}
\end{eqnarray}
where for simplicity we denote $\sigma_{1(2)}\equiv \sigma_{NV(SC)}$ hereafter. The last term is the beam-splitter form interaction,  which means photon hopping between the optical and microwave cavities.
\begin{figure}
\includegraphics[scale=0.35]{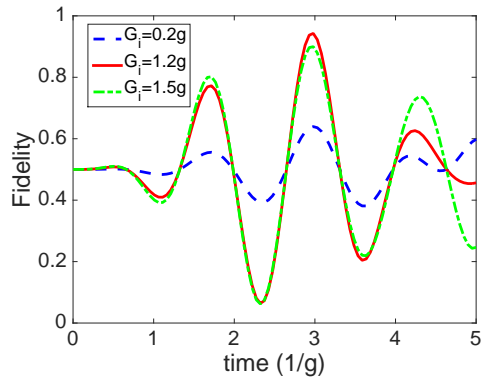}
\caption{\label{fig2} Fidelity as a function of time in the case of $g_{1}=g_{2}=g$ and $\theta = \pi/4$ with different $G_{i}$. The decoherence rates of them are chosen as $\kappa_{1}=0.1 g$ and $ \gamma_{1}=\kappa_{2}=\gamma_{2}=0.01 g$.}
\end{figure}

We now study  quantum state transfer from the NV center to the superconducting qubit in the zero- and one-excitation subspaces. Then the space can be spanned by the basis of $\{ |10\rangle_{12}|00\rangle_{ab},  |01\rangle_{12}|00\rangle_{ab},  |00\rangle_{12}|10\rangle_{ab},  |00\rangle_{12}|01\rangle_{ab}\} $. In our simulation we assume the initial state is $|\psi\rangle_{i}$ = (cos$\theta |0\rangle_{1}$+sin$\theta |1\rangle_{1})|0\rangle_{2}|00\rangle_{ab}$. Then the information encoded in the NV state is transferred to the superconducting qubit if we obtain a final state $|\psi\rangle_{f}$ = $|0\rangle_{1}$(cos$\theta |0\rangle_{2}$+sin$\theta |1\rangle_{2})|00\rangle_{ab}$. We estimate the performance of our protocol using the conditional fidelity $F(t)=\langle \psi_{1}|\rho_{sc}|\psi_{1}\rangle$, where $\psi_{1}$ is the target state to be transferred and $\rho_{sc}(t)$ is the reduced density matrix of the superconducting qubit at time $t$.

Due to the interaction of the system with its environment, we have to perform simulations by considering the decoherence of the NV center (described by a rate $\gamma_{1}$) and the superconducting qubit ($\gamma_{2}$), as well as the decay of the optical cavity ($\kappa_{1}$) and the microwave resonator ($\kappa_{2}$). With the above considerations we can simulate the system using the following Lindblad master equation \cite{QuantumOptics}:
 \begin{eqnarray}
\begin{split}
&\frac{d\rho}{dt}=-i[H_{transfer}, \rho]+\frac{1}{2}\kappa_{1}\zeta(\delta a)+\frac{1}{2}\gamma_{1}\zeta(\sigma^{z}_{1}) \\
&+\frac{1}{2}\kappa_{2}\zeta(b)+\frac{1}{2}\gamma_{2}\zeta(\sigma_{2})
\end{split}
\end{eqnarray}
where $\zeta(o)=2o\rho o^{\dagger}-o^{\dagger}o\rho-\rho o^{\dagger}o$ is the Lindblad operator for a given operator $o$. Note that the thermal occupations for microwave photons in the frequency range of GHz can be negligible at a temperature around 10 mK.

We focus on evaluating the performance of our scheme in the intermediate coupling regime ($G_{i} \sim g_{1}=g_{2}=g$), which we find is the best regime for quantum state transfer, as shown in an electromechanical coupling system \cite{QST_NVSC}. The transfer fidelity can be reduced as the electro-optical coupling strength deviates from the optimal case. From the time evolution of the fidelity as presented in Fig. \ref{fig2}, we find the fidelity can reach as high as 0.94 at time t=2.98 g under the given parameters. It indicates that the system can be reliable for high fidelity quantum information transfer. Moreover, conclusions from similar systems consisting of coupled resonators with the same form of the interaction Hamiltonian can be applied to our scheme \cite{QST_NVSC}.

%------------------------------------------------------------------------------------------------------------------------------------------------------
\section{Highly entangled states between the NV center and SC qubit}
In this section we consider the case when the driving laser is in the blue-sideband $\Delta=\omega_a-\omega_L=-\omega_{b}$. Then we can obtain the Hamiltonian under the rotating-wave approximation from Eq.(\ref{H_total}):
 \begin{eqnarray}
\begin{split}
&H_{entangle}=g_{1}(\delta a^{\dagger}\sigma_{1}+\delta a\sigma_{1}^{+})+g_{2}(b^{\dagger}\sigma_{2}+b\sigma_{2}^{+}) \\
&+G_{i}(\delta a^{\dagger} b^{\dagger}+\delta ab).
\end{split}
 \end{eqnarray}
The last term of the Hamiltonian describes the simultaneous creation or annihilation of a photon in the optical cavity and a photon in the microwave cavity. It's responsible for entangling the NV-cavity and SC qubit-resonator subsystems \cite{felicetti2014dynamical}.

 \begin{figure}
\includegraphics[scale=0.35]{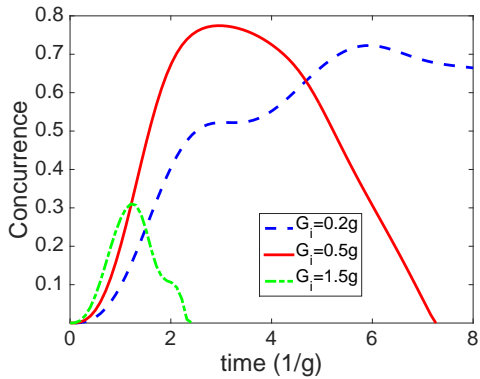}
\caption{\label{fig3} Concurrence as a function of time in the case of $g_{1}=g_{2}=g$ for different $G_{i}$. The decoherence rates are chosen as $\kappa_{1}=0.1g, \gamma_{1}=\kappa_{2}=\gamma_{2}=0.01g$.}
\end{figure}
% \begin{figure}
%\includegraphics[scale=0.45]{plot4.png}
%\caption{\label{fig4} Occupation in the optical cavity (red dashed line) and microwave resonator (black solid line) as a function of time. Here, $g_{1}=g_{2}=g$ and $G_{i}=1.2g$ and the other parameters are $\kappa_{1}=0.1g, \gamma_{1}=0.05g$ and $\kappa_{2}=\gamma_{2}=0.01g$.   }
%\end{figure}

We evaluate the entanglement of the NV center and the superconducting qubit by exploiting the $\emph{concurrence}$ C \cite{PhysRevLett.80.2245}. For the entanglement of formation of a mixed state $\rho$ of two qubits, the concurrence reads $C(\rho)=max\{0,\lambda_{1}-\lambda_{2}-\lambda_{3}-\lambda_{4}\}$, where $\lambda_{i}$ are the eigenvalues of the Hermitian matrix $R \equiv \sqrt{\sqrt{\rho}\tilde{\rho}\sqrt{\rho}}$ in decreasing order.  Here the spin-flipped state is $\tilde{\rho}=(\sigma_{y}\otimes \sigma_{y})\rho^{\ast}(\sigma_{y}\otimes \sigma_{y})$. Note that C goes from 0 to 1 as the state goes from an unentangled pure state such as $|10\rangle_{12}$ to a maximal entangled state such as $\frac{1}{\sqrt{2}}(|10 \rangle_{12}-|01\rangle_{12})$.
Again, we can simulate our system and get the time evolution of the concurrence through solving the following master equation numerically \cite{QuantumOptics}
 \begin{eqnarray}
\begin{split}
&\frac{d\rho}{dt}=-i[H_{entangle}, \rho]+\frac{1}{2}\kappa_{1}\zeta(\delta a)+\frac{1}{2}\gamma_{1}\zeta(\sigma^{z}_{1}) \\
&+\frac{1}{2}\kappa_{2}\zeta(b)+\frac{1}{2}\gamma_{2}\zeta(\sigma_{2}).
\end{split}
\end{eqnarray}

We assume that both the two qubits and the cavity and resonator are in their ground states at t=0. In the ideal case where there is no decay,  the maximum concurrence can reach 0.94 when $G_{i}=0.2g_{1}=0.2g_{2}$.  As shown in Fig. \ref{fig3}, the concurrence evolution is presented when considering the decoherence processes. Under the given parameters that $g_{1}=g_{2}=g$ and $G_{i}=0.5$ g, we find that the maximum of the concurrence is 0.77 at the time t=2.92 g. However, as one can imagine intuitively, the entanglement can decrease when $G_{i}$ has an offset with the optimal case. Therefore, a moderate electro-optic coupling is more favorable for achieving better entanglement of qubits. This conclusion is in line with that in the electromechanical system \cite{Entangle_NVSC}.For the above optimal case where $G_{i}=0.5$ g, the average photon numbers in the optical cavity and microwave resonator can reach 6 and 12 respectively.

\begin{figure}
\includegraphics[scale=0.35]{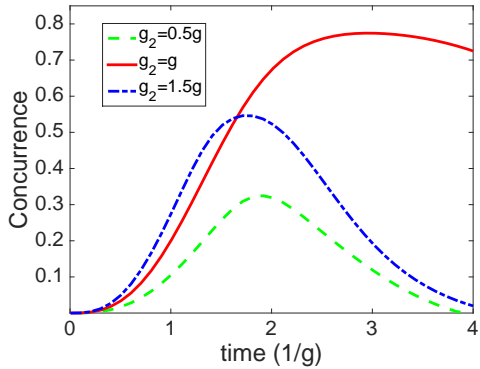}
\caption{\label{fig4} Concurrence as a function of time for different $g_{2}$ in the case of $g_{1}=g$ and $G_{i}=0.5g$. The decoherence rates are are chosen as $\kappa_{1}=0.1g, \gamma_{1}=\kappa_{2}=\gamma_{2}=0.01g$.}
\end{figure}

\begin{figure}
\includegraphics[scale=0.35]{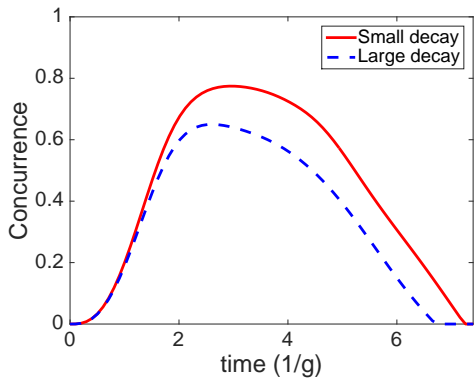}
\caption{\label{fig5} Concurrence as a function of time in the case of $g_{1}=g_{2}=g, G_{i}=0.5g$. The small decay curve corresponds the parameters that $\kappa_{1}=0.1g$, $\gamma_{1}=\kappa_{2}=\gamma_{2}=0.01g$, while the large decay curve has $\kappa_{1}=0.3g, \gamma_{1}=\kappa_{2}=\gamma_{2}=0.03g$.}
\end{figure}

Next we evaluate the effect of asymmetric coupling strength $g_{1}$ and $g_{2}$. In Fig. \ref{fig4} we present the time evolution of concurrence when $g_{1}\not=g_{2}$. Since the values of concurrence decrease as $g_{2}$ deviates away from $g_{1}$, the condition of $g_{1}=g_{2}$ should be satisfied in order to get better entanglement. In Fig. \ref{fig5}, we consider the concurrence evolution with different decoherence parameters. In the case of large decoherence of NV centers and superconducting quantum circuits ($\gamma_{1}=\gamma_{2}=0.03 g$), and large decay rates of the cavity and resonator ($\kappa_{1}=0.3 g, \kappa_{2}=0.03 g$), we observe that the maximum of concurrence is 0.65 at the time t=2.62 g. The simulation indicates that the time to reach the maximal entanglement doesn't vary significantly with the decoherence rates, and it's straightforward to see that we need to reduce the decay rates to obtain a higher concurrence value.

From the discussions above, we have shown that to get higher entanglement, the condition $g_{1}=g_{2}=g$ and $G_{i}\sim 0.5g$ should be satisfied and smaller decay rates are more favorable.

%---------------------------------------------------------------------------------------------------------------------------------------
\section{Implementation}

We now consider the experimental feasibility of the proposal. As illustrated above, one can couple an NV center with a WGM cavity. Strong coupling between an individual NV center and the fundamental WGM in a microsphere or microdisk has been reached and the coupling strength can be $g_{1}/2\pi \sim$ 0.3-1 GHz \cite{10.1021/nl061342r,10.1021/nl8032944,10.1021/nl8023627,PhysRevA.83.054306}. The state of the art Q factor for the WGM cavity can reach above $10^{10}$ and can be further enhanced  (up to $10^{12}$) by introducing the slow-light effect \cite{PhysRevLett.116.133902}. Then the photon loss rate can be as low as $\kappa_{1}/2\pi \sim 0.5$ MHz.With the help of adiabatic elimination of the excited level in NV $\Lambda-$level system, we can take advantage of the NV center's long coherence time, so here $\gamma_{1}$ can be very small.

On the other hand, microwave resonators with Q factors over $10^{6}$ and frequency  around 6 GHz has been reached which results in a decay rate of $\kappa_{2}/2\pi\sim3.5$ kHz \cite{apl_MWresonator}. The superconducting qubit can be protected by a transmission-line resonator. Energy relaxation time up to 44 $\mu s$ has been reported for a planar transmon qubit \cite{PhysRevLett.111.080502}, which leads to a decay rate $\gamma_{2}/2\pi \sim 3.5$ kHz. The strong \cite{10.1038/nature02851} and ultrastrong \cite{10.1038/nphys1730} coupling between the superconducting qubit and the transmission-line resonator has been experimentally demonstrated. Their coupling strength can be controlled very well using a flux-biased rf SQUID \cite{PhysRevLett.104.177004}. Therefore, in our scheme we assume $g_{2}$ here can be well tuned in a large range.

Finally, we discuss the strength of coupling between the NV-cavity and SC-resonator subsystems. One can utilize the electro-optic coefficient $n^{3}r \sim 300$ pm/V in lithium niobate \cite{QuantumElectronics,PhysRevA.94.053815} and d can be about 10 $\mu m$. Assuming $l/(c\tau) \sim 0.5$ and $C \sim 1$ pF, we can get $g_{i}/2\pi$ as large as 5 kHz. If we further consider the $|\alpha_{a}|$ term in the linearization process, the coupling strength can be increased by several orders, thus leading $G_{i} $ on the order of MHz and then $G_{i}$ can be comparable with coupling strengths $g_{1}$ and $g_{2}$. Therefore, the parameters in this proposal are in line with current experiment techniques.

\section{Conclusion}

In conclusion, in this work we have presented a proposal for efficiently transferring quantum states and generating entangled states between a single NV center and a superconducting qubit
in a hybrid electro-optic system. In this set-up, the coupling between the two subsystems, a single NV center and an optical cavity, as well as a superconducting qubit and a microwave cavity is via the electro-optic effect. By adjusting the relative parameters, we can achieve  high fidelity quantum state transfer by means of beam-splitter interactions, and generate highly entangled states between the NV spin and the superconducting qubit through two-mode squeezing interactions. This protocol may provide interesting applications in quantum information processing and quantum communication with single NV centers and superconducting qubits.

\section*{Acknowledgments}

This work is supported by the NSFC under
Grant Nos. 11774285 and 11474227. and the Fundamental
Research Funds for the Central Universities.

%\bibliography{resub_coulingNV_SC}
%\bibliographystyle{apsrev4-1}
%

\end{document}